\documentclass[oneside,english]{amsart}
\usepackage[T1]{fontenc}
\usepackage[latin1]{inputenc}
\usepackage{geometry}
\usepackage{graphicx}
\usepackage{amssymb}

\makeatletter
 \theoremstyle{plain}    
 \newtheorem{thm}{Theorem}[section]
 \numberwithin{equation}{section} 
 \numberwithin{figure}{section} 
 \theoremstyle{plain}
 \theoremstyle{remark}    
 \newtheorem*{acknowledgement*}{Acknowledgement} 

\DeclareMathOperator*{\fprod}{\overline{\prod}}

\usepackage{babel}
\makeatother
\begin{document}

\title{Finiteness and Dual Variables For Lorentzian Spin Foam Models}

\address{Department of Applied Mathematics, University of Western Ontario,
London, Ontario, Canada}

\email{jcherrin@uwo.ca}

\author{J. Wade Cherrington}

\begin{abstract}
We describe here some new results concerning the Lorentzian Barrett-Crane
model, a well-known spin foam formulation of quantum gravity. Generalizing
an existing finiteness result, we provide a concise proof of finiteness
of the partition function associated to all non-degenerate triangulations
of 4-manifolds and for a class of degenerate triangulations not previously
shown. This is accomplished by a suitable re-factoring and re-ordering
of integration, through which a large set of variables can be eliminated.
The resulting formulation can be interpreted as a {}``dual variables''
model that uses hyperboloid variables associated to spin foam edges
in place of representation variables associated to faces. We outline
how this method may also be useful for numerical computations, which
have so far proven to be very challenging for Lorentzian spin foam
models.
\end{abstract}
\maketitle

\section{Introduction}

Spin foam models offer a promising new quantum mechanical picture
of space-time geometry. The spin foam program has been developed as
a path integral formulation of loop quantum gravity%
\footnote{The reader is referred to \cite{key-38} for a recent review of loop
quantum gravity, and \cite{key-13} for a review of the spin foam
program specifically. %
}. Of the spin foam models proposed in the literature, perhaps the
most studied have been the Barrett-Crane models. Originally constructed
using representations of the group $Spin(4)$ associated to manifolds
of Riemannian signature \cite{key-9}, a subsequent model given by
Barrett and Crane in \cite{key-2} based upon representations of the
Lorentz group $SO(3,1)$ is considered to be more physically realistic.

While important strides have been made in numerical computations within
the Riemannian framework \cite{key-18,key-19,key-10}, numerical computations
with the Lorentzian model have proven much more difficult. An important
difference between the models lies in the space of representations,
which in the Lorentzian case is continuous rather than discrete. In
addition, the homogeneous space used to define vertex amplitudes is
non-compact in the Lorentzian case. 

The main result of the present work is that the Lorentzian partition
function can, by a process we shall refer to as {}``face factoring'',
be rearranged so that integration over the representation variables
is performed exactly. After applying this transformation, computational
or analytic effort can focus entirely on the homogeneous space integrals
and the one-dimensional integrals associated to each edge of the spin
foam. 

We note that the model that results from face factoring is strongly
in the spirit of the \emph{dual variables picture} proposed by Pfeiffer
in \cite{key-1}. While \cite{key-1} dealt specifically with a Riemannian
version of the Barrett-Crane model, our face-factored model can be
understood as the realization of a dual variables picture in the Lorentzian
case.

The organization of the paper is as follows. In the next section,
we recall the form of the Lorentzian Barrett-Crane amplitude and describe
the face factoring method. Included in this section is an illustration
of the method applied to the partition function of a simple 2-complex.
In Section 3, we prove that the partition function associated to triangulations
of a certain type is absolutely convergent in the face factored formulation,
which justifies the interchange of integration required by the method.
It follows that the original partition function is finite for triangulations
of this type, which includes both the non-degenerate triangulations
proven finite in \cite{key-4} as well as certain degenerate triangulations
that have not previously been proven finite. It should be noted here
that in the present work and in \cite{key-4}, finiteness is shown
for a specific choice of edge and face amplitudes; throughout this
work our main interest is in the choice%
\footnote{The physical content of these amplitude choices remains controversial;
for example, see \cite{key-14} for a critical discussion of this
issue. In Section 3, we examine how certain changes in the face amplitude
affect the type of degenerate triangulations for which we can show
finiteness.%
} due to Perez and Rovelli \cite{key-41}. We also show how the method
can be applied to an explicitly causal version of the Barrett-Crane
model due to Livine and Oriti \cite{key-3} for certain types of triangulations.
In Section 4, we briefly discuss the potential for numerical applications
of the method and end with some concluding remarks. A bound required
in Section 3 is proven in Appendix A, and in Appendix B we show that
the finiteness results of Section 3 (in which a closed 4-manifold
is assumed for clarity) also hold in the case of a 4-manifold with
boundary.

\section{Description of the Face Factoring Method}

\subsection*{Review of the Lorentzian Barrett-Crane Model}

To begin our discussion, we recall the form of a Lorentzian spin foam
model. Spin foam models are defined on simplicial 2-complexes constructed
from a set of vertices $V$, a set of edges $E$, and a set of polygonal
faces $F$. Because our present work is with the Barrett-Crane model,
we restrict ourselves to those 2-complexes which are the \emph{dual
2-skeleton} of a 4-manifold triangulation%
\footnote{In Section 3, we extend the strict definition of triangulation to
include degenerate triangulations which can be associated to topological
4-manifolds. %
}. The dual 2-skeleton is formed as follows: for every 4-simplex there
is an associated dual vertex, and any two dual vertices whose associated
4-simplices meet at a tetrahedron are connected by a dual edge. For
every face of the triangulation, the closed loop of 4-simplices that
share that face gives rise to a two-dimensional polygon in the dual
2-skeleton, the vertices of which are dual to the 4-simplices and
the edges of which are dual to the tetrahedra that are shared between
neighboring 4-simplices. An important property of this construction
is that the dual polygonal faces are in one-to-one correspondence
with the faces of the original 4-manifold triangulation. Further background,
including illustrations of dual skeleta, can be found in \cite{key-42}.

We now introduce the Lorentzian Barrett-Crane model, which assigns
to each dual 2-complex $\Delta$ a partition function $\mathcal{Z}_{\Delta}$
as follows: {\small \begin{equation}
\mathcal{Z}_{\Delta}=\underbrace{\int_{0}^{\infty}\cdots\int_{0}^{\infty}}_{f\in F}\left(\prod_{f\in F}\mathfrak{\mathcal{A}}_{f}\right)\left(\prod_{e\in E}\mathcal{A}_{e}\right)\left(\prod_{v\in V}\mathcal{A}_{v}\right)\prod_{f\in F}p_{\! f}^{2}\: dp_{f},\end{equation}
}where factors of the form $p_{\! f}^{2}$ arise from the measure
on the principal series of $SO(3,1)$ representations; the amplitudes
$\mathfrak{\mathcal{A}}_{f}$, $\mathfrak{\mathcal{A}}_{e}$, and
$\mathfrak{\mathcal{A}}_{v}$ will be defined shortly. For a given
2-complex, the state space associated with this partition function
is the product space of representation variables $p_{f}$ (one for
each face) and the multiple integration over this space is thought
of as a path integration over all spin configurations. To find the
transition amplitude between two spin networks (the basis states of
3-geometry in loop quantum gravity), a sum of $\mathcal{Z}_{\Delta}$
over all 2-complexes interpolating between the spin networks is required.
In the sum over 2-complexes, the amplitude $\mathcal{Z}_{\Delta}$
for each 2-complex will be weighted by an appropriate measure to provide
a normalization and possibly to regulate which 2-complexes can contribute.
The choice of measure and questions of convergence have yet to be
resolved --- progress is likely to depend on what results can be found
for the partition function of individual 2-complexes, which is the
focus of our present work.

In the expressions to follow, we write $\overline{\prod}$ to indicate
a formal product of symbols such as integral signs or measures. When
a relation such as $f\ni v$ appears below a product symbol (multiplicative
or formal), the product is taken over all of the objects on the left
hand side of relation that satisfy the relation; for example, in the
$f\ni v$ case the product would be over all faces $f$ such that
$v$ is a member of $f$. The symbols $v$, $e$, and $f$ will always
denote members of the sets $V$, $E$, and $F$, respectively.

We now turn to the definition of the amplitudes $A_{v}$, $A_{e}$,
and $A_{f}$ that appear in (2.1). Up to a regularization that will
be defined below, the Lorentzian Barrett-Crane vertex amplitude of
\cite{key-2} is defined as\begin{equation}
\mathcal{A}_{v}(p_{f})=\left(\overline{\prod_{e\ni v}}\int_{H_{+}^{3}}dx_{e}\right)\prod_{f\ni v}K_{p_{f}}(x_{e_{+}(f,v)},x_{e_{-}(f,v)}),\end{equation}
where the kernel function $K_{p_{f}}$ is \begin{equation}
K_{p_{f}}(x,y)=\frac{\sin(p_{f}\,\phi(x,y))}{p_{f}\sinh(\phi(x,y))}.\end{equation}
 The kernels $K_{p_{f}}$ are functions of the variables $x_{e_{-}(f,v)},x_{e_{+}(f,v)}\in H_{+}^{3}$
associated to the two dual edges that both contain $v$ and are contained
in $f$; the + and $-$ serve to distinguish the two tetrahedra within
a 4-simplex that share a face, but otherwise have no significance
--- any fixed convention can be chosen. As there are ten faces in
the 4-simplex to which a vertex is dual, the integrand is a product
of ten such kernels. The kernel function $K_{p_{f}}$ depends on the
hyperboloid variables through their hyperbolic distance $\phi$ on
$H_{+}^{3}$, which is defined as\begin{equation}
\phi(x,y)=\cosh^{-1}(x\cdot y)\end{equation}
for $x,y\in H_{+}^{3}$, where $H_{+}^{3}\equiv\{ x\in\mathbb{R}^{4}\mid x\cdot x=1,\: x_{0}>0\}$;
here $x\cdot y$ is the Minkowski inner product $x\cdot y\equiv x_{0}y_{0}-x_{1}y_{1}-x_{2}y_{2}-x_{3}y_{3}$. 

Although the expression for the vertex amplitude given above is generally
infinite, we shall regularize it by fixing the value of one $H_{+}^{3}$
variable and dropping the corresponding integral over $H_{+}^{3}$
--- the answer is independent of the choices \cite{key-2}.

It should be observed that the future 3-hyperboloid $H_{+}^{3}$ is
a homogeneous space of $SL(2,\mathbb{C})$ and, unlike the homogeneous
space $S^{3}$ that arises in the Riemannian model, it is non-compact.
The other basic difference with the Riemannian model is that the relevant
representations%
\footnote{In this we follow the original Lorentzian model of Barrett and Crane
and use exclusively the $(0,p)$ representations from the principal
series. While some proposals \cite{key-6} have been made which also
include half integer labelled representations $(k,0)$, most work
has been with the original proposal.%
} of $SL(2,\mathbb{C})$ are indexed by a continuous parameter $p_{f}$;
in place of a summation for each face the Lorentzian model requires
an integral. 

While the original model of \cite{key-2} specifies the vertex amplitude
$\mathcal{A}_{v}$ as given in (2.2), it leaves unspecified the edge
and face amplitudes $\mathcal{A}_{e}$ and $\mathcal{A}_{f}$. A proposal
introduced by Perez and Rovelli in \cite{key-41} and considered in
further work \cite{key-4,key-6} chooses%
\footnote{Some authors absorb the $p_{\! f}^{2}$ factors from the measure into
the definition of $\mathfrak{\mathcal{A}}_{f}(p_{f})$; in the present
work we keep the measure and face amplitude distinct.%
} $\mathcal{A}_{f}=1$ and $\mathcal{A}_{e}=\Theta_{4}(p_{1},p_{2},p_{3},p_{4})$
for edges in the interior of the 2-complex. The function $\Theta_{4}(p_{1},p_{2},p_{3},p_{4})$
is known as the \emph{eye diagram} and can be defined as follows:
\begin{equation}
\Theta_{4}(p_{1},p_{2},p_{3},p_{4})=\frac{2}{\pi p_{1}p_{2}p_{3}p_{4}}\int_{0}^{\infty}\frac{\sin(p_{1}r_{e})\sin(p_{2}r_{e})\sin(p_{3}r_{e})\sin(p_{4}r_{e})}{\sinh^{2}(r_{e})}\, dr_{e},\end{equation}
where $p_{1},\ldots,p_{4}$ are the spin variables labelling the four
faces containing $e$.

The results we have obtained with the face factoring technique have
primarily been for the amplitudes $\mathcal{A}_{f}=1$ and $\mathcal{A}_{e}=\Theta_{4}(p_{1},p_{2},p_{3},p_{4})$;
we shall henceforth refer to this choice as the \emph{Perez-Rovelli}
\emph{Model}.

\subsection*{The Face Factoring Method}

The major difficulty in computing $\mathcal{Z}_{\Delta}$ as presented
in (2.1) relates to the computation of the vertex amplitude (2.2).
For each choice of the ten $p_{f}$ variables, calculating a single
vertex amplitude requires extensive computational effort in the form
of Monte Carlo or quasi Monte Carlo integration. In the absence of
an exact expression or some other significantly more efficient means
of calculation, it is prohibitively expensive computationally to evaluate
the partition function for even the simplest 2-complexes. This situation
has provided the main motivation for finding an alternative approach
to calculating $\mathcal{Z}_{\Delta}$, in which no vertex amplitudes
are explicitly computed.

The idea of the present work is to perform the integration over the
spin variables $p_{f}$ first, leaving a formulation in which the
states to be integrated over are the configurations of $H_{+}^{3}$
variables originating with the vertex amplitude integrand and the
$r_{e}$ variables originating with the edge amplitude integrand.
This reverses the conventional integration order implicit in (2.1)
that has been adopted in most computational approaches to date. For
example, in computations for the Riemannian case \cite{key-18} the
vertex amplitudes (equal to integrals over the homogeneous space)
are computed for each spin configuration and the results summed to
give the partition function.

Given the difficulty in evaluating a Lorentzian vertex amplitude for
a single choice of spins, it is a rather fortuitous result that all
the spin dependence can be integrated out --- indeed, as we shall
see shortly, one obtains exact expressions in terms of the remaining
$H_{+}^{3}$ and $r_{e}$ variables. 

We define the method explicitly as follows. Assuming that the original
partition function is not affected by interchanging the order of integration,
we can write: \begin{equation}
\mathcal{Z}_{\Delta}=\left(\overline{\prod_{f\in F}}\int_{0}^{\infty}dp_{f}\right)\left(\prod_{f\in F}p_{\! f}^{2}\,\mathcal{A}_{f}(p_{f})\right)\left(\prod_{e\in E}\mathcal{A}_{e}(p_{f})\right)\end{equation}

\[
\left(\overline{\prod_{v\in V}}\left(\fprod_{e\ni v,e\neq e_{0}^{v}}\int_{H_{+}^{3}}dx_{(e,v)}\right)\left(\prod_{f\ni v}K_{p_{f}}(x_{e_{+}(f,v)},x_{e_{-}(f,v)})\right)\right)\]

\[
=\left(\overline{\prod_{v\in V}}\left(\fprod_{e\ni v,e\neq e_{0}^{v}}\int_{H_{+}^{3}}dx_{(e,v)}\right)\right)\left(\overline{\prod_{f\in F}}\int_{0}^{\infty}dp_{f}\right)\left(\prod_{f\in F}p_{\! f}^{2}\, A_{f}(p_{f})\right)\left(\prod_{e\in E}A_{e}(p_{f})\right)\]

\[
\prod_{v\in V}\left(\prod_{f\ni v}K_{p_{f}}(x_{e_{+}(f,v)},x_{e_{-}(f,v)})\right).\]
On the RHS of (2.6), all dependence on the $p_{f}$ variables is integrated
out before performing the integrals over $H_{+}^{3}$. As called for
by the regularization, one integration with respect to $H_{+}^{3}$
has been dropped from each vertex (the edge at $v$ for which integration
is dropped is denoted by $e_{0}^{v}$) . To perform integration over
the $p_{f}$ variables first, we consider the product of all the edge
and vertex amplitude integrands, and organize this product into factors
that each depend only upon a single $p_{f}$ variable. 

If we consider for the moment only the factors contributed by the
vertex amplitudes $\mathcal{A}_{v}$ given in (2.2), we note the following.
Due to the form of the integrand of $\mathcal{A}_{v}$ in terms of
the functions (2.3), we see that for any given face $f$ of the 2-complex
all of the factors involving the spin $p_{f}$ can be grouped together
into a \emph{vertex face factor} $F_{f}^{V}$ of the form\begin{equation}
F_{f}^{V}(p_{f},\phi_{i}^{f})=\frac{\sin(p_{f}\,\phi_{1}^{f})\sin(p_{f}\,\phi_{2}^{f})\cdots\sin(p_{f}\,\phi_{\mathrm{deg_{V}}(f)}^{f})}{p_{f}^{\mathrm{deg_{V}}(f)}},\end{equation}
where the $\phi_{i}^{f}$ denote distances on $H_{+}^{3}$ and $\mathrm{deg_{V}}(f)$
denotes the number of vertices contained in the dual face $f$. With
regard to the subscript $i$ indexing the $\phi_{i}^{f}$ variables,
we assume that a numbering of vertices in every dual face $f$ has
been chosen; the $i$ subscript runs over the values of the indexing
map $N_{f}(v)$ and the corresponding vertex $v$ is found by inverting
$N_{f}(v)$. Explicitly, we choose for each $f$ a bijective map of
the form $N_{f}:V_{f}\rightarrow\{1,\ldots,\mathrm{deg_{V}}(f)\}$,
where $V_{f}$ is the set of vertices in $f$. Although we have suppressed
the arguments of each $\phi_{i}^{f}$ in the formula above, for a
given $\phi_{i}^{f}$ letting $v=N_{f}^{-1}(i)$ we see that $\phi_{i}^{f}=\phi^{f}(x_{e_{-}(f,v)},x_{e_{+}(f,v)})$
is a function of the $H_{+}^{3}$ variables $x_{e_{-}(f,v)}$ and
$x_{e_{+}(f,v)}$.

Although the face amplitude for the Perez-Rovelli model has no spin
dependence ($A_{f}=1)$, the edge amplitude (2.5) is a non-trivial
function of the four spins variables associated to an edge of the
2-complex. While the eye diagram can be integrated to an exact expression
in terms of the product of spin variables and the hyperbolic cotangent
function \cite{key-2}, the result does not have a form in which the
spins $p_{f}$ appear in separate factors. However, we observe from
(2.5) that \emph{before} integration with respect to $r_{e}$, the
$p_{f}$ dependent part of the integrand is a product of factors of
the form $\frac{\sin(p_{f}r_{e})}{p_{f}}$. Hence, if we collect factors
from the different edge amplitude integrands that depend on the same
$p_{f}$, we have a product for each face which has the same form
as the vertex face factor (2.7), but with the $r_{e}$ variables playing
the role of the hyperbolic distance variables $\phi_{i}^{f}$. Therefore,
if the $r_{e}$ variables are integrated only after integrating with
respect to the $p_{f}$, we can define an overall face factor that
takes into account contributions from both edge and vertex amplitude
kernels. Our final form for the face factors are then\begin{equation}
F_{f}(p_{f},\phi_{i}^{f},r_{e})=\frac{\sin(p_{f}\,\phi_{1}^{f})\cdots\sin(p_{f}\,\phi_{\deg_{V}(f)}^{f})\sin(p_{f}\, r_{e(f,1)})\cdots\sin(p_{f}\, r_{e(f,\deg_{E}(f))})}{p_{f}^{\deg_{V}(f)+\deg_{E}(f)-2}},\end{equation}
where $\deg_{E}(f)$ denotes the number of edges contained in the
face $f$ and the $e(f,i)$ selects the $i$th edge%
\footnote{As with the face-vertex numbering $N_{f}(v)$, we assume that for
every face a numbering of its edges $M(f,e)$ has been chosen. We
use an edge valued subscript for the $r_{e}$ as there is only one
such variable for every edge. %
} contained in the face $f$. It should be noted that we have absorbed
the $p_{\! f}^{2}$ terms from the measure into our definition of
$F_{f}$ by lowering the power of $p_{f}$ in the denominator by two. 

Having defined our face factors, we can now rewrite the partition
function in terms of the integrals of the face factors $F_{f}(p_{f},\phi_{i}^{f},r_{e})$
and products of $\mathrm{sinh}$ functions that depend only upon the
$r_{e}$ and $\phi_{i}^{f}$: \begin{equation}
\mathcal{Z}_{\Delta}=\left(\overline{\prod_{e}}\int_{0}^{\infty}dr_{e}\right)\left(\overline{\prod_{v}}\left(\fprod_{e\ni v,e\neq e_{0}^{v}}\int_{H_{+}^{3}}dx_{(e,v)}\right)\right)\left(\prod_{e}\frac{2}{\pi\sinh^{2}(r_{e})}\right)\end{equation}

\[
\left(\prod_{v}\left(\prod_{f\ni v}\frac{1}{\sinh(\phi_{N_{f}(v)}^{f})}\right)\right)\left(\prod_{f}\int_{0}^{\infty}F_{f}(p_{f},\phi_{i}^{f},r_{e})\, dp_{f}\right).\]
Formula (2.9) is our explicit \emph{face factored formulation} for
the partition function of the Perez-Rovelli model; in Section 3 we
shall prove it is equal to the original model for all non-degenerate
triangulations and certain degenerate triangulations. 

From (2.9), we see that face factoring requires interchanging the
order of integration among the $p_{f}$, the $r_{e}$, and the hyperboloid
integrals --- all of which are improper. In order to justify the interchange,
it is sufficient that at least one of the orderings of integration
is absolutely integrable. This follows from the Tonelli-Hobson test,
a corollary of the well-known Fubini theorem in the case of integrals
on an unbounded domain; see for example \cite{key-39}. In Section
3 below, we justify this interchange for a general class of triangulated
4-manifolds by establishing absolute integrability for the face factored
formulation of the partition function. 

We consider next an illustration of the face-factoring method for
the 2-complex dual to a particular triangulation of the 4-sphere $S^{4}$.

\subsection*{Example face factoring of a Barrett-Crane model with trivial edge
amplitude}

For simplicity, we consider in this section a modified Perez-Rovelli
model with (2.5) replaced by a trivial edge amplitude so that\[
\mathcal{A}_{f}=1,\quad\mathcal{A}_{e}=1.\]
The standard Barrett-Crane vertex amplitude\[
\mathcal{A}_{v}=\left(\fprod_{e\ni v,e\neq e_{0}^{v}}\int_{H_{+}^{3}}dx_{e}\right)\left(\prod_{f\ni v}K_{p_{f}}(x_{e_{+}(f,v)},x_{e_{-}(f,v)})\right)\]
is used; recall from (2.3) that the kernel functions are\[
K_{p_{f}}(x,y)=\frac{\sin(p_{f}\,\phi(x,y))}{p_{f}\sinh(\phi(x,y))}.\]

Consider the triangulation of $S^{4}$ resulting from taking the boundary
of the 5-simplex. In this triangulation, there are 6 four-simplices,
15 tetrahedra, and 20 triangles. We shall work with the dual 2-skeleton
of this triangulation, in which there are 6 vertices, 15 edges, and
20 polygonal faces; the relevant face factors $F_{f}(p_{f},\phi_{i}^{f})$
will be deduced from counting and symmetry arguments. Because each
vertex is contained in ten faces, there will be ten $K_{p_{f}}(x,y)$
factors per vertex for a total of sixty $K_{p_{f}}(x,y)$ factors.
In our present case, all faces have an associated face factor of the
same form by symmetry. Since by definition there is a face factor
for each face $f$ and there are twenty faces for the entire 2-complex,
we see that each face factor has the form \begin{equation}
F_{f}(p_{f},\phi_{i}^{f})=\frac{\sin(p_{f}\phi_{1}^{f})\sin(p_{f}\phi_{2}^{f})\sin(p_{f}\phi_{3}^{f})}{p_{f}},\end{equation}
where the three $\phi_{i}^{f}$ are associated to the three vertices
in $f$. Having found the face factor, we consider the integral that
appears in the face-factored form of the partition function definition
given in (2.9):\begin{equation}
\int_{0}^{\infty}F_{f}(p_{f},\phi_{i}^{f})\, dp_{f}=\int_{0}^{\infty}\frac{\sin(p_{f}\phi_{1}^{f})\sin(p_{f}\phi_{2}^{f})\sin(p_{f}\phi_{3}^{f})}{p_{f}}\, dp_{f}\end{equation}

\[
=\frac{1}{8\pi}(\mathrm{sign}(-\phi_{1}^{f}+\phi_{2}^{f}+\phi_{3}^{f})-\mathrm{sign}(-\phi_{1}^{f}+\phi_{2}^{f}-\phi_{3}^{f})-\mathrm{sign}(\phi_{1}^{f}+\phi_{2}^{f}+\phi_{3}^{f})+\mathrm{sign}(\phi_{1}^{f}+\phi_{2}^{f}-\phi_{3}^{f})).\]
As suggested above, this integral is not difficult to evaluate to
an exact closed form in terms of the distance variables $\phi_{i}^{f}$.
We note here that given the difficulty of evaluating a single Lorentzian
vertex amplitude, it may be surprising that integrating over all the
spin variables can be done in such a way that the result is a product
of simple functions of the $H_{+}^{3}$ variables. 

We again emphasize that to equate the face factored form of (2.9)
with the original formulation of the partition function, interchanging
the order of the improper integrals needs to be justified. We won't
provide such a justification for this example case --- however, in
Section 3 we prove that the face factoring method is valid for the
actual case of interest, the Lorentzian Barrett-Crane model with the
Perez-Rovelli face and edge amplitudes. Before doing so, we shall
give an exact closed form expression for the face factor integrations
that arise in the Perez-Rovelli model.

\subsection*{Closed form for the Perez-Rovelli face factor integrations.}

In the Perez-Rovelli model, the most general face factor is given
by (2.8), and the integrals required to eliminate dependence on $p_{f}$
for a given $f$ are of the form\[
\int_{0}^{\infty}\left(\prod_{v\in f}\frac{\mathrm{sin}(p_{f}\phi_{N_{f}(v)}^{f})}{p_{f}}\right)\left(\prod_{e\in f}\frac{\mathrm{sin}(p_{f}r_{e})}{p_{f}}\right)p_{\! f}^{2}\, dp_{f}.\]
Clearly, this integral can be expressed using a product of $\mathrm{sinc}$
functions as

\[
\left(\prod_{v\in f}\phi_{N_{f}(v)}^{f}\right)\left(\prod_{e\in f}r_{e}\right)\int_{0}^{\infty}\left(\prod_{v\in f}\mathrm{sinc}(p_{f}\phi_{N_{f}(v)}^{f})\right)\left(\prod_{e\in f}\mathrm{sinc}(p_{f}r_{e})\right)p_{\! f}^{2}\, dp_{f}.\]
Were it not for the factor of $p_{\! f}^{2}$ due to the measure,
one would be integrating a product of $\mathrm{sinc}$ functions with
arguments scaled by the $\phi_{i}^{f}$ and $r_{e}$ variables. If
this were the case, a rather remarkable closed form given by Borwein
et. al. in \cite{key-45} could be directly applied. This formula,
which will be useful for our actual case as well, we give here as
follows.

Without loss of generality, let an arbitrary face $f$ in the 2-complex
be chosen. We define a new set of variables $d_{i}$ so that $d_{i}=\phi_{i+1}^{f}$
for $0\leq i<\mathrm{deg_{V}}(f)$ and $d_{i}=r_{e(f,i+1-\mathrm{deg_{V}}(f))}$
for $\mathrm{deg_{V}}(f)\leq i<\mathrm{deg_{V}}(f)+\mathrm{deg_{E}}(f)$.
As the $d_{i}$ are all non-negative, we can apply from \cite{key-45}
the result:\begin{equation}
\int_{0}^{\infty}\left(\prod_{v\in f}\frac{\mathrm{sin}(p_{f}\phi_{N_{f}(v)}^{f})}{p_{f}}\right)\left(\prod_{e\in f}\frac{\mathrm{sin}(p_{f}r_{e})}{p_{f}}\right)\, dp_{f}\end{equation}

\[
=\int_{0}^{\infty}\left(\prod_{i=0}^{n}\frac{\mathrm{sin}(p_{f}d_{i})}{p_{f}}\right)\, dp_{f}=\frac{\pi}{2}\frac{1}{2^{n}n!}\sum_{\gamma\in\{-1,1\}^{n}}\epsilon_{\gamma}b_{\gamma}^{n}\,\mathrm{sign}(b_{\gamma})\]
where $n=\mathrm{deg_{V}}(f)+\mathrm{deg_{E}}(f)-1$ and $\gamma=(\gamma_{1},\cdots,\gamma_{n})\in\{-1,1\}^{n}$.
The $b_{\gamma}$ and $\epsilon_{\gamma}$ are defined by\[
b_{\gamma}=d_{0}+\sum_{k=1}^{n}\gamma_{k}d_{k},\quad\epsilon_{\gamma}=\prod_{k=1}^{n}\gamma_{k}.\]
Remarkably, it turns out that this formula enables us to handle the
Perez-Rovelli case as well --- we simply differentiate both sides
of (2.12) twice with respect to (any) one of the $d_{i}$ parameters.
On the LHS the differentiation is passed under the integral sign;
choosing $d_{0}$ as our parameter to differentiate we have: \begin{equation}
-\frac{\partial^{2}}{\partial d_{0}^{2}}\int_{0}^{\infty}\left(\prod_{i=0}^{n}\frac{\mathrm{sin}(p_{f}d_{i})}{p_{f}}\right)\, dp_{f}=\int_{0}^{\infty}-\frac{\partial^{2}}{\partial d_{0}^{2}}\left(\prod_{i=0}^{n}\frac{\mathrm{sin}(p_{f}d_{i})}{p_{f}}\right)\, dp_{f}\end{equation}

\[
=\int_{0}^{\infty}\left(\prod_{i=0}^{n}\frac{\mathrm{sin}(p_{f}d_{i})}{p_{f}}\right)p_{\! f}^{2}\, dp_{f}=-\frac{\partial^{2}}{\partial d_{0}^{2}}\left(\frac{\pi}{2}\frac{1}{2^{n}n!}\sum_{\gamma\in\{-1,1\}^{n}}\epsilon_{\gamma}b_{\gamma}^{n}\,\mathrm{sign}(b_{\gamma})\right)\]
Note that differentiating $\sin(p_{f}d_{0})$ twice with respect to
$d_{0}$ gives the desired factor of $p_{\! f}^{2}$ multiplying the
negative of $\mathrm{sin}(p_{f}d_{0})$. Thus (2.13) provides a closed
form expression for all the face factor integrals that may appear
in the Perez-Rovelli model. In practice, one can also evaluate these
types of integrals using symbolic integration with software such as
Maple or Mathematica.

\section{Finiteness Results}

The finiteness of the Lorentzian Barrett-Crane partition function
with the Perez-Rovelli choice for face and edge amplitudes was an
important finding established%
\footnote{See also \cite{key-48} for further detail.%
} in \cite{key-4} for arbitrary \emph{non-degenerate} 2-complexes.
In this section, we prove that the partition function is absolutely
integrable for a more general class of 2-complexes. This proof both
justifies the interchange of improper integrals required by face factoring
and at the same time proves finiteness of the model for all non-degenerate
triangulations and a certain class of degenerate triangulations. We
characterize this class of degenerate triangulations and show how
it is sensitive to certain changes in the face amplitude. Note that
the proof we give in this section applies to \emph{closed} two-complexes.
The result also holds for 2-complexes with boundary; we describe in
Appendix B how the proof of the closed case is changed to account
for the presence of boundaries. 

We start by introducing an absolute bound on the Perez-Rovelli partition
function in its face-factored form (2.9) above.\begin{equation}
\left|\mathcal{Z}_{\Delta}\right|\leq\left(\overline{\prod_{e}}\int_{0}^{\infty}dr_{e}\right)\left(\overline{\prod_{v}}\left(\fprod_{e\ni v,e\neq e_{0}^{v}}\int_{H_{+}^{3}}dx_{(e,v)}\right)\right)\left(\prod_{e}\frac{2}{\pi\sinh^{2}(r_{e})}\right)\end{equation}

\[
\left(\prod_{v}\left(\prod_{f\ni v}\frac{1}{\sinh(\phi_{N_{f}(v)}^{f})}\right)\right)\left(\prod_{f}\int_{0}^{\infty}\left|F_{f}(p_{f},\phi_{i}^{f},r_{e})\right|\: dp_{f}\right).\]
Introducing the inequality (A.5) proven in Appendix A: \begin{equation}
\prod_{f}\left(\int_{0}^{\infty}\left|F(p_{f},\phi_{i}^{f},r_{e})\right|dp_{f}\right)\leq\prod_{f}\left(\frac{4}{3}\left(\prod_{i=1}^{\deg_{V}(f)}\phi_{i}^{f}\prod_{j=1}^{\deg_{E}(f)}r_{e(f,j)}\right)^{1-\frac{3}{\deg_{V}(f)+\deg_{E}(f)}}\right).\end{equation}
Hence, if the RHS of the bound\begin{equation}
\left|\mathcal{Z}_{\Delta}\right|\leq\left(\overline{\prod_{e}}\int_{0}^{\infty}dr_{e}\right)\left(\overline{\prod_{v}}\left(\fprod_{e\ni v,e\neq e_{0}^{v}}\int_{H_{+}^{3}}dx_{(e,v)}\right)\right)\left(\prod_{e}\frac{2}{\pi\sinh^{2}(r_{e})}\right)\end{equation}

\[
\left(\prod_{v}\left(\prod_{f\ni v}\frac{1}{\sinh(\phi_{N_{f}(v)}^{f})}\right)\right)\left(\prod_{f}\,\frac{4}{3}\left(\prod_{i=1}^{\deg_{V}(f)}\phi_{i}^{f}\prod_{j=1}^{\deg_{E}(f)}r_{e(f,j)}\right)^{1-\frac{3}{\deg_{V}(f)+\deg_{E}(f)}}\right)\]
is finite then our face factored form of the partition function (and
hence the original form) is finite. 

To integrate out the $r_{e}$ variables in our bound, for each $e$
we collect together all the $r_{e}$ dependence into a single term
of the form\begin{equation}
\int_{0}^{\infty}\frac{r_{e}^{\alpha_{e}}}{\sinh^{2}(r_{e})}\, dr_{e},\end{equation}
where $\alpha_{e}>0$ is the overall power of $r_{e}$ appearing in
(3.3) due to contributions from the four faces that contain $e$.

We claim that for $\alpha_{e}>1$, quantities of this form are always
finite. To see this we divide the integration as follows:\[
\int_{0}^{\infty}\frac{r_{e}^{\alpha_{e}}}{\sinh^{2}(r_{e})}\, dr_{e}=\int_{0}^{\frac{\ln2}{2}}\frac{r_{e}^{\alpha_{e}}}{\sinh^{2}(r_{e})}\, dr_{e}+\int_{\frac{\ln2}{2}}^{\infty}\frac{r_{e}^{\alpha_{e}}}{\sinh^{2}(r_{e})}\, dr_{e}.\]
 Consider the first term. Although the integrand is unbounded at $r_{e}=0$,
$\sinh(r_{e})\rightarrow r_{e}$ as $r_{e}\rightarrow0$, so near
zero the integrand behaves as $r^{-(2-\alpha_{e})}$, which is integrable
for $\alpha_{e}>1$, but will otherwise diverge. Because $\frac{e^{2x}}{16}<\sinh^{2}(x)$
for $x>\frac{\ln2}{2}$, we can bound the second term as\[
\int_{\frac{\ln2}{2}}^{\infty}\frac{r_{e}^{\alpha_{e}}}{\sinh^{2}(r_{e})}\, dr_{e}<16\int_{\frac{\ln2}{2}}^{\infty}r_{e}^{\alpha_{e}}e^{-2r_{e}}\, dr_{e}\]
which is finite, as the polynomial $r_{e}^{\alpha_{e}}$ is exponentially
damped as $r_{e}\rightarrow\infty$. Thus all factors involving $r_{e}$
can be integrated to give finite factors if for each edge $e$ the
inequality $\alpha_{e}>1$ holds. We will return below to a discussion
of this $\alpha_{e}>1$ condition and the constraints it places on
which degenerate triangulations can be shown finite.

Next we need to check that integration over the hyperboloid variables
always yields finite factors. Upon integrating out the $r_{e}$ as
described, we have reduced the bound (3.3) to: \begin{equation}
\left|\mathcal{Z}_{\Delta}\right|\leq C\,\left(\overline{\prod_{v}}\left(\fprod_{e\ni v,e\neq e_{0}^{v}}\int_{H_{+}^{3}}dx_{(e,v)}\right)\right)\left(\prod_{v}\left(\prod_{f\ni v}\frac{1}{\sinh(\phi_{N_{f}(v)}^{f})}\right)\right)\prod_{f}\left(\prod_{i=1}^{\deg_{V}(f)}\phi_{i}^{f}\right)^{1-\frac{3}{\deg_{V}(f)+\deg_{E}(f)}}\end{equation}

\[
=C\,\prod_{v}\left(\left(\fprod_{e\ni v,e\neq e_{0}^{v}}\int_{H_{+}^{3}}dx_{(e,v)}\right)\left(\prod_{f\ni v}\frac{1}{\sinh(\phi_{N_{f}(v)}^{f})}\right)\prod_{f\ni v}\left(\phi_{N_{f}(v)}^{f}\right)^{1-\frac{3}{\deg_{V}(f)+\deg_{E}(f)}}\right).\]
where the $r_{e}$ integrations and all other constants have been
absorbed into an overall constant $C$. 

From this expression, we see that to complete our proof we require
\begin{equation}
\left(\fprod_{e\ni v,e\neq e_{0}^{v}}\int_{H_{+}^{3}}dx_{(e,v)}\right)\prod_{f\ni v}\frac{\left(\phi_{N_{f}(v)}^{f}\right)^{1-\frac{3}{\deg_{V}(f)+\deg_{E}(f)}}}{\sinh(\phi_{N_{f}(v)}^{f})}\end{equation}
 for each $v$ to be finite. Note that these integrals are of the
same form of the vertex amplitude (2), but with modified kernel functions.
In recent work by Dan Christensen \cite{key-43}, a proof is given
in which a class of integrals that includes those of the form (3.6)
are shown to be finite. From this we can conclude that \emph{all hyperboloid
integrations produce finite factors}. 

Therefore, absolute integrability for our bound on the face factored
formulation depends only upon finiteness of the $r_{e}$ variable
integrations, for which we found above that $\alpha_{e}>1$ is a necessary
and sufficient condition.

We analyze the $\alpha_{e}>1$ condition as follows. For all non-degenerate
triangulations, $\deg_{V}(f)+\deg_{E}(f)\geq6$, hence the power of
$r_{e}$ appearing in each face factor is at least $(1-\frac{3}{6})=\frac{1}{2}$.
Since each $r_{e}$ is associated to a tetrahedron, it will appear
in 4 face factors. Hence $\alpha_{e}\geq(4\cdot\frac{1}{2})=2>1$,
and so we conclude that \emph{all non-degenerate triangulations are
absolutely integrable.} This justifies the face factoring method for
these triangulations and gives an alternative finiteness proof to
that of \cite{key-4}.

More generally, the condition on $\alpha_{e}$ in terms of the vertex
and edge degrees $\deg_{V}(f)$ and $\deg_{E}(f)$ can be given as
$\alpha_{e}=\sum_{f\ni e}\left(1-\frac{3}{\deg_{V}(f)+\deg_{E}(f)}\right)>1$.
If we consider degenerate triangulations where $\deg_{E}(f)\geq2$
and $\deg_{E}(f)\geq2$, this condition is equivalent to the following
statement:

\begin{thm}
\emph{The Perez-Rovelli partition function $\mathcal{Z}_{\Delta}$
is absolutely integrable if for every $e\in\Delta$,} $\deg_{V}(f)+\deg_{E}(f)>4$
\emph{for at least one} $f$ \emph{containing} $e$. 
\end{thm}
This is a generalization of the finiteness proof of \cite{key-4},
which was limited to non-degenerate triangulations. We shall see next
that some degenerate triangulations satisfy this condition and some
do not.

\subsection*{Finiteness for Degenerate Triangulations}

In the case where a dual 2-complex contains a face $f$ where $\deg_{V}(f)$
or $\deg_{E}(f)$ is less than 3, it is dual to a degenerate triangulation.
Such a {}``triangulation'' is not strictly speaking a triangulation
in the usual sense of a simplicial complex, as it contains 4-simplices
whose intersection contains more than one tetrahedra. While visualizing
4-dimensional geometry is challenging, we can draw lower-dimensional
analogs of degenerate triangulations; two examples are shown in Figure
3.1. 

Using Theorem 3.1, the 4-dimensional analog%
\footnote{In the analog, two four-simplices are glued along two of their tetrahedra.%
} of Figure 3.1(a) can be shown finite. Observe that although the interior
degenerate edges (indicated with thickened lines) have a face where
$\deg_{E}(f)=2$ and $\deg_{V}(f)=2$, they also have 3 faces where
$\deg_{E}(f)>2$ and $\deg_{V}(f)>2$, and so applying our criterion
above we see that \emph{$\mathcal{Z}_{\Delta}$} is finite. 

The four dimensional analog of Figure 3.1(b), in which two 4-simplices
intersect along all five of their boundary tetrahedra, results in
$\deg_{E}(f)=2$ and $\deg_{V}(f)=2$ for all faces inside the tetrahedron
of the intersection; for edges dual to these tetrahedra $\alpha_{e}=4\cdot\frac{1}{4}=1$
so integration of our face factor bound in the $r_{e}$ variable diverges.
Hence our proof fails to show that such 2-complexes are finite. We
remark here that our Theorem 3.1 is a sufficient condition but we
have not shown it to be necessary --- our work does not rule out finiteness
of the face factored form for this type of triangulation. However,
even if the face factored form can be shown \emph{not} to be absolutely
integrable, the original form of the partition function may still
be finite but only conditionally convergent. Currently, numerical
computations are underway by the author to explore some of these possibilities
for 2-complexes arising from these degenerate triangulations. 

A final observation relates to how possible changes in the face amplitude
affect the integrability of degenerate triangulations. While the edge
and vertex amplitudes we have used here are fairly well accepted and
can be given some physical motivation, it may be of interest to consider
alternatives to the $\mathcal{A}_{f}(p_{f})=1$ choice. If we allow
ourselves to consider a face amplitude of the form\[
\mathcal{A}_{f}(p_{f})=\frac{1}{p_{f}^{\gamma}}\]
for $\gamma\geq0$, we can construct a new bound (following the steps
of Appendix A) in which the convergence condition becomes $\alpha_{e}=\sum_{f\ni e}\left(1-\frac{3-\gamma}{\deg_{V}(f)+\deg_{E}(f)}\right)>1$.
The analog of Theorem 3.1 is than the requirement that\[
\mathrm{deg_{V}}(f)+\mathrm{deg_{E}}(f)+\frac{4}{3}\gamma>4\]
for some $f$ containing $e$. 

It is interesting to observe that for $\gamma>0$, this criterion
is still met even if $\deg_{V}(f)=\deg_{E}(f)=2$ for \emph{all} the
faces of a tetrahedron, the case which fails for the original amplitude
$\mathcal{A}_{f}(p_{f})=1$ corresponding to $\gamma=0$. Hence, with
the original model such 2-complexes give a bound that is divergent
but at the {}``margin'' of integrability --- any non-zero $\gamma$
results in a model where all degenerate triangulations satisfying
$\deg_{E}(f)\geq2$ and $\deg_{E}(f)\geq2$ are finite.

\begin{figure}[t]
\includegraphics[%
  scale=0.69]{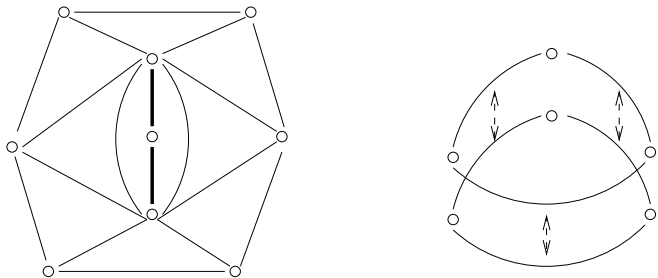}

\caption{(a) A degenerate triangulation of the two-dimensional disc. (b) A
degenerate triangulation of the 2-sphere. Dotted arrows indicate edge
identifications. }
\end{figure}

\section{Applications}

\subsection*{Generalization to the Livine-Oriti Causal Barrett-Crane Model}

In the Livine-Oriti causal model developed in \cite{key-3}, the kernel
functions for the vertex amplitude $\mathcal{A}_{v}$ are modified
to take into account face orientations $\varepsilon_{v}^{f}\in\{-1,1\}$
relative to a vertex $v$ that arise when imposing causality. The
causal kernel replacing the original kernel (2.3) is given as\begin{equation}
K_{p_{f}}(x,y)=\frac{e^{i\varepsilon_{v}^{f}p_{f}\phi_{N_{f}(v)}^{f}(x,y)}}{p_{f}\sinh(\phi_{N_{f}(v)}^{f}(x,y))}.\end{equation}
From the form of this kernel, we can write the vertex face factors
as\begin{equation}
F_{f}^{V}(p_{f},\phi_{i}^{f})=\frac{1}{p_{f}^{\mathrm{deg_{V}}(f)}}\prod_{v\in f}e^{i\varepsilon_{v}^{f}p_{f}\phi_{N_{f}(v)}^{f}}.\end{equation}
The integral $\int_{0}^{\infty}F_{f}^{V}(p_{f},\phi_{i}^{f})\: dp_{f}$
diverges in the general case, because the numerator approaches a constant
as $p_{f}\rightarrow0$ while the denominator approaches zero at least
linearly; this gives a divergent integral for $\mathrm{deg_{v}}(f)\geq1$,
which applies to any triangulation. Hence, contributions to the overall
face factor from face and edge amplitudes must be relied upon to give
a finite face factored partition function.

\subsubsection*{Convergence criterion for the face amplitude:}

In \cite{key-3}, the edge amplitude remains the eye diagram function
when passing to the causal case. Taking the causal vertex and edge
amplitudes as fixed, we show in this section how the face amplitude
has to be carefully tuned to avoid divergence of the face factored
formulation. Let us consider the full causal face factor with the
edge amplitude kernels and a face amplitude of the form $\mathcal{A}_{f}(p_{f})=p_{\! f}^{\beta}$
for a real number $\beta$:\begin{equation}
F_{f}(p_{f},\phi_{i}^{f},r_{e})=\frac{1}{p_{f}^{\mathrm{deg_{V}}(f)+\mathrm{deg_{E}(f)}-2-\beta}}\left(\prod_{v\in f}e^{i\varepsilon_{v}^{f}p_{f}\phi_{N_{f}(v)}^{f}}\right)\left(\prod_{e\in f}\sin(p_{f}r_{e})\right).\end{equation}
 Observing that the product of $\mathrm{deg_{E}(f)}$ sine functions
can cancel against the $p_{f}^{\mathrm{deg_{E}}(f)}$ at $p_{f}=0$,
to avoid divergence we arrive at the condition\[
\mathrm{deg_{V}}(f)-3<\beta.\]
In addition, if $\beta\geq\mathrm{deg_{V}}(f)+\mathrm{deg_{E}(f)}-2$
then $F_{f}(p_{f},\phi_{i}^{f},r_{e})$ for general $\phi$ fails
to have a limit as $p_{f}\rightarrow\infty$, and so its integral
is not defined. Hence a necessary finiteness condition for a given
$\beta$ is \begin{equation}
\mathrm{deg_{V}}(f)-1<\beta+2<\mathrm{deg_{V}}(f)+\mathrm{deg_{E}(f)}.\end{equation}

For $\beta=0$ as has been proposed, (4.4) becomes $\mathrm{deg_{V}}(f)$$\leq2$
and $\mathrm{deg_{V}}(f)+\mathrm{deg_{E}(f)}>2$; thus we see that
the former condition restricts us to degenerate triangulations. For
$\beta\geq1$ however, degenerate triangulations are completely ruled
out. Note here that in (4.4) the parameter $\beta$ is bounded from
both above and below.

We stress here that failure of convergence for the face factored partition
function does not imply divergence of the original form; it may still
be that the original partition function is conditionally integrable,
precluding the identification of the two forms. 

To summarize, we find that for the Oriti-Livine model there are some
rather strong constraints on $\mathcal{A}_{f}(p_{f})$ necessary for
the existence of a finite face factored formulation. An interesting
goal for future work is to determine whether the model with original
integration ordering escapes these constraints by being conditionally
convergent, or if it in fact is divergent in the same cases as are
divergent in the face factored formulation.

\subsection*{Numerical Applications}

Within the context of Monte Carlo methods applied directly to the
partition function, the face factored formulation is advantageous
in several respects. The integrations eliminated involved a highly
oscillatory product of sine functions; the integration of these functions
leads to much smoother functions of the remaining variables. In fact,
it can be shown that the remaining functions are always non-negative
\cite{key-40}. The features of smoothing and positivity are both
desirable in Monte Carlo methods for improving the accuracy achievable
for a given number of samples. Currently, computations are underway
to evaluate the partition function for sufficiently small 2-complexes
where the number of integration dimensions is still not prohibitive.

Apart from direct Monte Carlo methods which have severe scaling limitations,
the face factored form of the partition function may be amenable to
statistical mechanical methods similar to those used in lattice quantum
field theory computations. In such computations, one avoids calculating
the partition function itself, but rather generates a sequence of
samples in configuration space which has the same distribution as
that determined by the partition function. In this way, expectation
values of observables can be found to good approximation without calculating
the entire partition function. The Metropolis algorithm is one such
method that was successfully applied \cite{key-18} in calculating
expectation values for the Riemannian Barrett-Crane model. 

Many statistical mechanical methods require that the partition function
amplitude%
\footnote{The integrand of the partition function $\mathcal{Z}_{\Delta}$; a
function of $p_{f}$ variables in the original formulation and a function
of $H_{+}^{3}$ and $r_{e}$ variables associated to tetrahedra in
the face factored formulation.%
} be non-negative for all configurations (i.e. $\!$the partition function
amplitude is the exponential of some real-valued action). Clearly,
before face factoring the Perez-Rovelli amplitude as a function of
all the spin variables fails to have such positivity due to its definition
in terms of sine functions. However, after the $p_{f}$ variables
have been integrated, it is rather remarkable that for closed spin
foams the amplitude as a function of tetrahedral variables is always
non-negative \cite{key-40}. The potential for applying lattice field
theory techniques to spin foam models using {}``dual'' tetrahedral
variables has been emphasized by Pfeiffer in \cite{key-1}, and may
open the door to computing with much larger 2-complexes. 

On a cautionary note, the physical interpretation of a model defined
purely in terms of tetrahedral variables seems ambiguous at this time.
Based upon connections to the Regge action in certain limits and general
geometric considerations, it has been argued%
\footnote{For instance in \cite{key-3}; a dual variable framework is also developed
in some detail for the Euclidean case in \cite{key-1}.%
} that the hyperboloid variables can be interpreted as the time-like
normals to the tetrahedra (which are purely spacelike by the Barrett-Crane
construction). While this viewpoint is certainly promising, it is
unclear to the author how these interpretations can be related back
to the canonical approach.

\section{Conclusions}

In this paper we have shown how the Lorentzian Barrett-Crane spin
foam model can be reformulated so that integration with respect to
the spin variables $p_{f}$ is performed before all other integrations
to yield exact expressions in terms of tetrahedral variables. Absolute
integrability was proven for the face factored formulation, justifying
the interchange of integration and establishing finiteness for a large
class of triangulations. This class includes all non-degenerate triangulations
and a certain class of degenerate triangulations, extending the finiteness
proof of \cite{key-4}. In addition to the face-factoring transformation,
an essential ingredient of this proof is a recent finiteness result
due to Christensen \cite{key-43} for a large class of 10j-like integrals.

We have also described how the type of degenerate triangulations proven
finite by our method depends upon the choice of face amplitude. Given
this result, it may be worthwhile to investigate more thoroughly the
physical significance of the degenerate triangulations. While the
method can also be applied to the causal Livine-Oriti model, we found
that given their choice of edge and vertex amplitude, there is a rather
stringent constraint between the form of the face amplitude and the
types of triangulations that allow a finite face factoring formulation.

An interesting feature of the Barrett-Crane model revealed by our
work is that edge and vertex amplitudes contribute kernels to the
face factors of the same form; this is no accident and can be traced
back to the diagrammatics used to derive the model \cite{key-2}.
Hence, an obvious question is whether or not generalizations or modifications
of the Barrett-Crane model may still allow a face factoring formulation.
Specifically, one could consider whether the Barrett-Crane type model
proposed in \cite{key-6}, with mixed representations, may be susceptible
to a face factoring treatment. Moreover, it may be interesting to
investigate whether any analog of face factoring is possible for spin
foam models not of the Barrett-Crane type. 

\begin{acknowledgement*}
The author would like to thank Dan Christensen, Igor Khavkine and
Joshua Willis for very useful discussions and suggestions. The author
also acknowledges the financial support of NSERC and OGS. 
\end{acknowledgement*}
\appendix

\section{Product Bound for General Face Factor Integrations}

In this appendix we find functions of $r_{e}$ and $\phi_{i}^{f}$
that bound from above the possible integrated face factors that arise
in the Perez-Rovelli model. The bounding functions have the simple
form of a product of $r_{e}$ and $\phi_{i}^{f}$ variables raised
to a power that depends only upon the number of edges and vertices
that the face contains. These bounding functions are used in Section
3 to show the absolute integrability of the face-factored formulation.

To begin, we recall that the most general face factor that can arise
for the Perez-Rovelli model has the form

\begin{equation}
F_{f}(p_{f},\phi_{i}^{f},r_{e})=\frac{\sin(p_{f}\phi_{1}^{f})\cdots\sin(p_{f}\phi_{\deg_{V}(f)}^{f})\sin(p_{f}r_{e(f,1)})\cdots\sin(p_{f}r_{e(f,\deg_{E}(f))})}{p_{f}^{\deg_{V}(f)+\deg_{E}(f)-2}},\end{equation}
where the $r_{e}$ and $\phi_{i}^{f}$ are non-negative real numbers
coming from edge and vertex amplitudes, respectively. To see this,
note simply that in both the edge and vertex amplitude each sine function
in the numerator is accompanied by a factor of $p_{f}$ in the denominator.
The measure then contributes a factor of $p_{\! f}^{2}$, reducing
the degree of the denominator by 2. Based on this form, we can immediately
establish two types of bounds:\begin{equation}
\left|F_{f}(p_{f},\phi_{i}^{f},r_{e})\right|\leq\frac{p_{f}\phi_{1}^{f}\cdots p_{f}\phi_{\deg_{V}(f)}^{f}p_{f}r_{e(f,1)}\cdots p_{f}r_{e(f,\deg_{E}(f))}}{p_{f}^{\deg_{V}(f)+\deg_{E}(f)-2}}=\left(\prod_{i=1}^{\deg_{V}(f)}\phi_{i}^{f}\prod_{j=1}^{\deg_{E}(f)}r_{e(f,j)}\right)p_{\! f}^{2}\end{equation}
and

\begin{equation}
\left|F_{f}(p_{f},\phi_{i}^{f},r_{e})\right|=\frac{\left|\sin(p_{f}\phi_{1}^{f})\cdots\sin(p_{f}\phi_{\deg_{V}(f)}^{f})\sin(p_{f}r_{e(f,1)})\cdots\sin(p_{f}r_{e(f,\deg_{E}(f))})\right|}{p_{f}^{\deg_{V}(f)+\deg_{E}(f)-2}}\end{equation}

\[
\leq\frac{1}{p_{f}^{\deg_{V}(f)+\deg_{E}(f)-2}}.\]

The first inequality (A.2) is valid for $p_{f}\geq0$, while the second
bound (A.3) can only be used for $p_{f}>0$. Integrating both sides
of these inequalities we can form the following estimate for the integrated
face factor:

\begin{equation}
\int_{0}^{\infty}\left|F_{f}(p_{f},\phi_{i}^{f},r_{e})\right|dp_{f}\leq\int_{0}^{M}\left(\prod_{i=1}^{\deg_{V}(f)}\phi_{i}^{f}\prod_{j=1}^{\deg_{E}(f)}r_{e(f,j)}\right)p_{\! f}^{2}\, dp_{f}+\int_{M}^{\infty}\frac{1}{p_{f}^{\deg_{V}(f)+\deg_{E}(f)-2}}\end{equation}

\[
=\frac{1}{3}\left(\prod_{i=1}^{\deg_{V}(f)}\phi_{i}^{f}\prod_{j=1}^{\deg_{E}(f)}r_{e(f,j)}\right)M^{3}+\frac{1}{M^{\deg_{V}(f)+\deg_{E}(f)-3}},\]
for any $M>0$. Since we are integrating $p_{f}$ for a given set
of the $\phi_{i}^{f}$ and $r_{e}$, we can make this $M$ a function
$M(\phi_{i}^{f},r_{e})$ of these variables. We make the following
choice:

\[
M(\phi_{i}^{f},r_{e})=\left(\prod_{i=1}^{\deg_{V}(f)}\phi_{i}^{f}\prod_{j=1}^{\deg_{E}(f)}r_{e(f,j)}\right)^{-\frac{1}{\deg_{V}(f)+\deg_{E}(f)}},\]
which upon substituting into (A.4) gives:

\begin{equation}
\int_{0}^{\infty}\left|F_{f}(p_{f},\phi_{i}^{f},r_{e})\right|dp_{f}\leq\frac{1}{3}\left(\prod_{i=1}^{\deg_{V}(f)}\phi_{i}^{f}\prod_{j=1}^{\deg_{E}(f)}r_{e(f,j)}\right)M^{3}+\frac{1}{M^{\deg_{V}(f)+\deg_{E}(f)-3}}\end{equation}

\[
=\frac{1}{3}\left(\prod_{i=1}^{\deg_{V}(f)}\phi_{i}^{f}\prod_{j=1}^{\deg_{E}(f)}r_{e(f,j)}\right)^{1-\frac{3}{\deg_{V}(f)+\deg_{E}(f)}}+\left(\prod_{i=1}^{\deg_{V}(f)}\phi_{i}^{f}\prod_{j=1}^{\deg_{E}(f)}r_{e(f,j)}\right)^{1-\frac{3}{\deg_{V}(f)+\deg_{E}(f)}}\]

\[
=\frac{4}{3}\left(\prod_{i=1}^{\deg_{V}(f)}\phi_{i}^{f}\prod_{j=1}^{\deg_{E}(f)}r_{e(f,j)}\right)^{1-\frac{3}{\deg_{V}(f)+\deg_{E}(f)}}.\]
Although we assume $M>0$ for this derivation to be valid, if any
of the $\phi_{i}^{f}$ or $r_{e}$ are zero than the original integrand
(A.1) and hence the integral over $p_{f}$ vanishes; hence the bound
(A.5) is valid for any choice of $\phi_{i}^{f}$ and $r_{e}$. As
it relates to the discussion at the end of Section 3, we mention here
that for a face amplitude of the form $\mathcal{A}_{f}(p_{f})=\frac{1}{p_{f}^{\gamma}}$
the above proof can be generalized to arrive at a bound:

\[
\int_{0}^{\infty}\left|F_{f}(p_{f},\phi_{i}^{f},r_{e})\right|dp_{f}\leq\left(1+\frac{1}{3-\gamma}\right)\left(\prod_{i=1}^{\deg_{V}(f)}\phi_{i}^{f}\prod_{j=1}^{\deg_{E}(f)}r_{e(f,j)}\right)^{1-\frac{3-\gamma}{\deg_{V}(f)+\deg_{E}(f)}}\]
by following essentially the same steps.

\section{Finiteness for spin foams with boundary }

As a path integral theory of quantum gravity, a spin foam model defines
a transition amplitude from one 3-geometry to another for a given
{}``history'' --- a triangulated 4-manifold whose boundary is the
union of the ingoing and outgoing 3-geometries. 

In the case where the 2-complex is dual to a triangulated 4-manifold
with boundary, a straightforward extension of our finiteness proof
for the closed case can be given. The spin variables associated to
each face in the boundary are taken as boundary data, so these are
not integrated in defining the partition function. As well, for dual
edges and faces coming from boundary tetrahedra and triangles, one
needs to take the square root of the edge and face amplitudes so that
the partition function multiplies when two histories are {}``glued''
along a common boundary. We shall denote the sets of edges and faces
dual to the boundary tetrahedra and triangles as $\partial E$ and
$\partial F$, respectively. Rather than working directly with the
square root of the edge amplitude $\Theta_{4}(p_{1},\ldots,p_{4})$,
in this proof we use the following bound:\begin{equation}
\sqrt{\Theta_{4}(p_{1},\ldots,p_{4})}<1+\Theta_{4}(p_{1},\ldots,p_{4}).\end{equation}
Applying this bound to the partition function of a 2-complex $\Delta$
with boundary we have\begin{equation}
\mathcal{Z}_{\Delta}=\underbrace{\int_{0}^{\infty}\cdots\int_{0}^{\infty}}_{f\notin\partial F}\left(\prod_{e\in\partial E}\sqrt{\Theta_{4}(p_{1},\ldots,p_{4})}\right)\left(\prod_{e\notin\partial E}\Theta_{4}(p_{1},\ldots,p_{4})\right)\left(\prod_{v\in V}\mathcal{A}_{v}(p_{1},\ldots,p_{10})\right)\prod_{f\notin\partial F}p_{\! f}^{2}\: dp_{f}\end{equation}

\[
<\underbrace{\int_{0}^{\infty}\cdots\int_{0}^{\infty}}_{f\notin\partial F}\left(\prod_{e\in\partial E}1+\Theta_{4}(p_{1},\ldots,p_{4})\right)\left(\prod_{e\notin\partial E}\Theta_{4}(p_{1},\ldots,p_{4})\right)\left(\prod_{v\in V}\mathcal{A}_{v}(p_{1},\ldots,p_{10})\right)\prod_{f\notin\partial F}p_{\! f}^{2}\: dp_{f}.\]
Considering the RHS of (B.2), we see that expanding the product of
$1+\Theta_{4}$ functions results in a sum of partition functions
in which boundary tetrahedra are either trivial ($\mathcal{A}_{e}=1)$
or have the eye diagram amplitude $\mathcal{A}_{e}=\Theta_{4}(p_{1},\ldots,p_{4})$.
We will show that any partition function of this form is finite.

First we handle the integrals with respect to the edge variables $r_{e}$.
For the given 2-complex $\Delta$, let a partition function be chosen
where the amplitude for any boundary edge is either $\mathcal{A}_{e}=1$
or $\mathcal{A}_{e}=\Theta_{4}(p_{1},\ldots,p_{4})$, and all other
amplitudes agree with the Perez-Rovelli model. We can extend the face
factoring method to a partition functions of this form as follows.
We begin by defining a boundary face factor $F_{f}^{\partial}$ associated
to any face on the boundary as:\begin{equation}
F_{f}^{\partial}(p_{f},\phi_{i}^{f},r_{e})=\frac{\sin(p_{f}\phi_{1}^{f})\sin(p_{f}\phi_{2}^{f})\cdots\sin(p_{f}\phi_{\deg_{V}(f)}^{f})\sin(p_{f}r_{1})\sin(p_{f}r_{2})\cdots\sin(p_{f}r_{\deg_{E}^{\Theta}(f)})}{p_{f}^{\deg_{V}(f)+\deg_{E}^{\Theta}(f)}}\end{equation}
 where the function $\deg_{E}^{\Theta}(f)$ counts the boundary edges
of type $\Theta_{4}$ contained in $f$. Observe that since the $p_{f}$
are fixed on the boundary rather than integrated over, the measure
factor of $p_{\! f}^{2}$ does not enter into the definition of $F_{f}^{\partial}(p_{f},\phi_{i}^{f},r_{e})$;
the functions $F_{f}^{\partial}(p_{f},\phi_{i}^{f},r_{e})$ and not
their integrals with respect to $p_{f}$ multiply into the partition
function directly. The analog of the inequality (A.2) for faces in
the boundary can be given as \begin{equation}
\left|F_{f}^{\partial}(p_{f},\phi_{i}^{f},r_{e})\right|\leq\frac{p_{f}\phi_{1}^{f}p_{f}\phi_{2}^{f}\cdots p_{f}\phi_{\deg_{V}(f)}^{f}p_{f}r_{1}p_{f}r_{2}\cdots p_{f}r_{\deg_{E}^{\Theta}(f)}}{p_{f}^{\deg_{V}(f)+\deg_{E}^{\Theta}(f)}}=\left(\prod_{i=1}^{\deg_{V}(f)}\phi_{i}^{f}\prod_{j=1}^{\deg_{E}^{\Theta}(f)}r_{e(f,j)}\right).\end{equation}
In the open spin foam case, the inequality (3.2) is modified to include
the boundary face factors:\[
\prod_{f\notin\partial\Delta}\int_{0}^{\infty}\left|F_{f}(p_{f},\phi_{i}^{f},r_{e})\right|\, dp_{f}\prod_{f\in\partial\Delta}\left|F_{f}^{\partial}(p_{f},\phi_{i}^{f},r_{e})\right|\]

\[
\leq\prod_{f\notin\partial\Delta}\left(\frac{4}{3}\prod_{i=1}^{\deg_{V}(f)}\phi_{i}^{f}\prod_{j=1}^{\deg_{E}(f)}r_{e(f,j)}\right)^{1-\frac{3}{\deg_{V}(f)+\deg_{E}(f)}}\prod_{f\in\partial\Delta}\left(\prod_{i=1}^{\deg_{V}(f)}\phi_{i}^{f}\prod_{j=1}^{\deg_{E}^{\Theta}(f)}r_{e(f,j)}\right).\]
For tetrahedra in the interior, finiteness goes through as in the
closed case. For tetrahedra on the boundary, there is an integration
over $r_{e}$ for any edge $e$ of type $\Theta_{4}$; for such edges
one collects all the $r_{e}$ dependence to give factors of the form:\[
\int_{0}^{\infty}\frac{r_{e}^{4}}{\sinh^{2}(r_{e})}\, dr_{e}.\]
As this integral is clearly finite, integrating with respect to the
$r_{e}$ for all $e$ in the interior and boundary always yields finite
factors. For boundary tetrahedra where $\mathcal{A}_{e}=1$, there
is no $r_{e}$ variable to be integrated so finiteness of the bound
is not affected. 

The final step is to show that the result of integrating the partition
function bound over the hyperboloid variables is always finite. For
a 4-simplex which has all of its faces on the interior, we recall
from the proof of the closed case that integration with respect to
hyperboloid variables factors into products of the form\[
\left(\fprod_{e\ni v,e\neq e_{0}^{v}}\int_{H_{+}^{3}}dx_{(e,v)}\right)\prod_{f\ni v}\frac{\left(\phi_{N_{f}(v)}^{f}\right)^{1-\frac{3}{\deg_{V}(f)+\deg_{E}(f)}}}{\sinh(\phi_{N_{f}(v)}^{f})},\]
each of which can shown finite by invoking the result of \cite{key-43}.
For 4-simplices that have one or more faces in the boundary, one instead
has\begin{equation}
\left(\fprod_{e\ni v,e\neq e_{0}^{v}}\int_{H_{+}^{3}}dx_{(e,v)}\right)\prod_{f\ni v,f\notin\partial\Delta}\frac{\left(\phi_{N_{f}(v)}^{f}\right)^{1-\frac{3}{\deg_{V}(f)+\deg_{E}(f)}}}{\sinh(\phi_{N_{f}(v)}^{f})}\prod_{f\ni v,f\in\partial\Delta}\frac{\phi_{N_{f}(v)}^{f}}{\sinh(\phi_{N_{f}(v)}^{f})},\end{equation}
which again are shown to be finite in \cite{key-43} --- thus we have
checked all hyperboloid integrations lead to finite factors. 

As our bound (B.2) is a finite sum of partition functions which are
finite, we conclude that the Perez-Rovelli Lorentzian partition function
is finite for all 2-complexes dual to triangulated 4-manifolds with
boundary, on the condition that the interior faces meet the same criteria
with respect to $\deg_{V}(f)$ and $\deg_{E}(f)$ as were found for
the closed case.

\end{document}